\newcommand{\vecj}{\mbox{\boldmath$j$}}

\newcommand{\vecp}{\mbox{\boldmath$p$}}

\newcommand{\vecr}{\mbox{\boldmath$r$}}

\newcommand{\vecv}{\mbox{\boldmath$v$}}

\newcommand{\vecA}{\mbox{\boldmath$A$}}
\newcommand{\vecB}{\mbox{\boldmath$B$}}

\newcommand{\vecE}{\mbox{\boldmath$E$}}

\newcommand{\vece}{\mbox{{\boldmath$\hat{e}$}}}

\newcommand{\dfd}{{\rm d}}

\newcommand{\half}{\frac{1}{2}}

\newcommand{\ie}{{\em i.e.}}

\documentclass[aps,prl,twocolumn,groupedaddress,showpacs]{revtex4}

\begin{document}


\title{Classical diamagnetism, magnetic interaction energies, and repulsive forces in magnetized plasmas}


\author{Hanno Ess\'en}
\email{hanno@mech.kth.se}
\affiliation{Department of Mechanics, KTH\\ SE-100 44 Stockholm,
Sweden}


\date{2011 April}

\begin{abstract}
The Bohr-van Leeuwen theorem is often summarized as saying that there is no classical magnetic susceptibility, in particular no diamagnetism. This is seriously misleading. The theorem assumes position dependent interactions but this is not required by classical physics. Since the work of Darwin in 1920 it has been known that the magnetism due to classical charged point particles can only be described by allowing velocity dependent interactions in the Lagrangian. Legendre transformation to an approximate Hamiltonian can give an estimate of the Darwin diamagnetism for a system of charged point particles. Comparison with experiment, however, requires knowledge of the number of classically behaving electrons in the sample. A new repulsive effective many-body force, which should be relevant in plasmas, is predicted by the Hamiltonian.
\end{abstract}


\maketitle


\section{Introduction}
The Bohr-van Leeuwen \cite{bohr,van_leeuwen,BKvanvleck,carati&benfenati&galgani} (BvL) theorem states that the magnetic susceptibility of a classical system of charged {\em point} particles interacting via some {\em position} dependent potential energy is zero. Further support for this theorem can be found in \cite{darwin3,berger,kaplan&mahanti,pradhan&seifert,saha&al}. Landau showed that quantum mechanics can explain diamagnetism in metals \cite{landauDiaMag}. Recent work, however, shows that perfect conductors exhibit classical perfect diamagnetism \cite{mahajan,fiolhais&al}, seemingly in blatant conflict with Bohr and van Leeuwen. The BvL theorem has also been questioned by Dubrovskii \cite{dubrovskii} on the grounds that it neglects a relevant constant of the motion other than the energy.  Usually the message of the BvL theorem is summarized as proving the nonexistence of classical diamagnetism \cite{kaplan&mahanti,pradhan&seifert,saha&al,bandy&datta,kumar&kumar}. We will show here that this is highly misleading by displaying an accurate energy expression for a system of charged point particles in an external magnetic field. We then also discuss the Hamiltonian corresponding to this energy and draw some general conclusions about the behavior of magnetized plasmas.

It is obvious that one can treat magnetic susceptibility using classical models if one gives up the assumption of {\em point} particles. Using classical objects that are extended balls of charge one can find a classical explanation of diamagnetism \cite{essen89}, and classical models with dipoles can explain paramagnetism. One can, however, reasonably argue that these are not fundamental in the same way that point monopole particles are. Since the work of Charles Galton Darwin \cite{darwin} in 1920 it has been known that the correct Lagrangian for a system of charged point particles requires {\em velocity} dependent interactions. Once the BvL assumption of only {\em position} dependent interactions is relaxed one finds classical diamagnetism effortlessly, as we now proceed to show.

\section{On the magnetic energy of a system of point charges}
The electromagnetic energy of a system can be expressed in many ways. Here we first assume that we are dealing with charged point particles in vacuum, \ie\ no dipoles. It is then sufficient to consider the two fields $\vecE$ and $\vecB$, or equivalently the potentials $\phi, \vecA$. The various expressions for the energy of a system, and their interrelations, that then can be written down have been reviewed by Franklin \cite{franklin07}. One well known expression for the energy is given by,
\begin{equation}\label{eq.energy1}
E = \sum_j \half  m_j \vecv_j^2 + \frac{1}{8\pi}\int (\vecE^2 + \vecB^2) \,\dfd V .
\end{equation}
The electric energy is however normally taken into account by finding the electrostatic potential, $\phi(\vecr_j; \vecr_k)$ at particle $j$, due to the other particles $k\neq j$ of the system. This gives,
\begin{equation}\label{eq.energy2}
E =  \sum_j \half \left[  m_j \vecv_j^2 + e_j \phi(\vecr_j; \vecr_k)\right]+ \frac{1}{8\pi}\int  \vecB^2 \,\dfd V ,
\end{equation}
where self-interactions are assumed removed. Consider now the magnetic energy. A moving charged particle produces the magnetic field,
\begin{equation}\label{eq.mag.field.part.j}
\vecB_j(\vecr) = \frac{e_j}{c} \frac{\vecv_j \times (\vecr - \vecr_j)}{|\vecr - \vecr_j|^3},
\end{equation}
to first order in $v/c$. The total (internal) field is then
\begin{equation}\label{eq.int.mag.field}
\vecB_{\rm i}(\vecr) =\sum_j \vecB_j (\vecr)
\end{equation}
To estimate the energy we should then introduce this in the integral and integrate over all of space. To get finite results one must again ignore self-interactions. This means that we put,
\begin{equation}\label{eq.Bi.sqared.no.self}
E_{\rm im} = \sum_{j<k} \frac{1}{4\pi} \int  \vecB_j \cdot \vecB_k \,\dfd V,
\end{equation}
for the internal magnetic interaction energy.
The calculation gives (Breitenberger \cite{breitenberger}),
\begin{equation}
E_{\rm im} =
\sum_{j<k} \frac{e_j e_k}{ r_{kj}} \frac{[\vecv_j \cdot\vecv_k +
(\vecv_j \cdot\vece_{kj})(\vecv_j\cdot\vece_{kj})]}{2c^2}.
\end{equation}
Here $r_{kj} = |\vecr_j - \vecr_k|$ and $\vece_{kj}=(\vecr_j - \vecr_k)/r_{kj}$. This is the Darwin magnetic energy expression that follows without approximation from the Darwin Lagrangian. By introducing the internal vector potential,
\begin{equation}\label{eq.darw.vec.pot}
\vecA_{\rm i}(\vecr_j;\vecr_k,\vecv_k)=\sum_{k\neq j} \frac{e_k}{r_{kj}}\frac{\vecv_k + (\vecv_k\cdot \vece_{kj})\vece_{kj}}{2c},
\end{equation}
one finds,
\begin{equation}\label{eq.int.mag.A}
E_{\rm im} =
\sum_{j} \frac{e_j}{2c} \vecv_j \cdot \vecA_{\rm i}(\vecr_j;\vecr_k,\vecv_k),
\end{equation}
as an alternative expression.

If there is also an external magnetic field, $\vecB_{\rm e}(\vecr)$, the magnetic energy will be,
\begin{equation}\label{eq.energy3}
 \frac{1}{8\pi}\int  (\vecB_{\rm i} + \vecB_{\rm e})^2 \,\dfd V \equiv E_{\rm im} + E_{\rm ie} + E_{\rm e}.
\end{equation}
This form of the magnetic energy makes it obvious that minimization occurs when the internal field is as much as possible of opposite direction and of equal magnitude compared to the external field; {\em hence diamagnetism.} To calculate it we note that doing the square gives three terms. The first of these give (\ref{eq.Bi.sqared.no.self}-\ref{eq.int.mag.A}) above. To calculate the other two terms we assume that the magnetic field is constant $\vecB_{\rm e} =  B_{\rm e} \vece_z$ in the region ($r<R$) where our system of charged particles resides. In order to get finite results we take the external field to be a dipole field outside some large enough radius $R$, so that for $r>R$ we have $\vecB_{\rm e}(\vecr) = (B_{\rm e} /2)[3(\vece_z \cdot \vece_r) \vece_r - \vece_z](R/r)^3 $. This gives us,
\begin{equation}\label{eq.mag.energy.int}
E_{\rm ie}= \frac{1}{4\pi}\int  \vecB_{\rm i} \cdot \vecB_{\rm e} \,\dfd V = \sum_j \frac{e_j}{c}\vecv_j \cdot \vecA_{\rm e}(\vecr_j) ,
\end{equation}
where,
\begin{equation}\label{eq.ext.vec.pot}
\vecA_{\rm e}(\vecr) =\frac{\theta(r)}{2}  B_{\rm e}\vece_z \times \vecr,
\end{equation}
with $\theta(r) = \min(1,R^3/r^3)$, is the vector potential of our external field \footnote{This is the field produced by current on a sphere of radius $R$ corresponding to a rigidly rotating constant surface charge density.}. Finally we have the constant energy of the external field, which is given by,
\begin{equation}\label{eq.mag.energy.ext}
E_{\rm e}= \frac{1}{8\pi}\int  \vecB_{\rm e}^2 \,\dfd V = \frac{R^3}{4} B_{\rm e}^2.
\end{equation}
Assuming a constant external field extending to infinity, as is often done in simplified treatments, makes not only $E_{\rm e}$, but also $E_{\rm ie}$, infinite.

Summarizing, we have obtained the energy,
\begin{eqnarray}
\nonumber
E(\vecr_k,\vecv_k) = \sum_j \half \left[  m_j \vecv_j^2 + e_j \phi(\vecr_j; \vecr_k)\right]+ \\
\label{eq.energy.final}    \\
\nonumber
 \sum_{j} \frac{e_j}{c} \vecv_j \cdot \left[\half \vecA_{\rm i}(\vecr_j;\vecr_k,\vecv_k) + \vecA_{\rm e}(\vecr_j) \right] + E_{\rm e}.
\end{eqnarray}
The BvL theorem is obtained by neglecting the internal magnetic field $\vecB_{\rm i}$, \ie\ the $\vecA_{\rm i}$-contribution here.

The Darwin Lagrangian ${\cal L}(\vecr_k,\vecv_k)$ is obtained by changing the sign of the electrostatic term in (\ref{eq.energy.final}).
It should be emphasized that the Darwin Lagrangian, which is presented in many advanced textbooks \cite{BKjackson3,BKlandau2,BKliboff,BKpage&adams,BKpodolsky,BKschwinger&al}, describes most of classical electromagnetism \cite{essen09}, except that radiation and highly relativistic effects are neglected \cite{essen07}. The perfect diamagnetism of perfect conductors and superconductors is well described in the Darwin formalism \cite{essen05,fiolhais&al} since these systems, due to the absence of dissipation, can be studied using classical electrodynamics.

\section{Legendre transform of the Darwin Lagrangian}
To investigate susceptibilities using statistical mechanics, however, requires the Hamiltonian. In principle the Hamiltonian ${\cal H}$ is trivially obtained from the Lagrangian ${\cal L}$ by means of the Legendre transform,
\begin{equation}\label{eq.legendre.transf}
{\cal H}(\vecr_k,\vecp_k) = \sum_{j=1}^N \vecp_j \cdot \vecv_j - {\cal L}(\vecr_k,\vecv_k),
\end{equation}
after having solved for the velocities $\vecv_j=\dot{\vecr}_j$ in terms of the momenta $\vecv_j(\vecr_k,\vecp_k)$, in the equations $\vecp_j=\partial {\cal L}/\partial \vecv_j$. When ${\cal L}$ is the Darwin Lagrangian this calculation turns out to be difficult. Only a few texts discuss the Darwin Hamiltonian and mostly a first order correction, called the Breit term in relativistic quantum mechanics, is arrived at. The first to seriously consider the problem beyond the first order approximation were Primakoff and Holstein \cite{primakoff}. They pointed out that when $v/c$ is not small there will be non-negligible effective many-body forces in the Hamiltonian formalism. Even if $v/c$ is small, however, the terms in question can be large when many particles contribute, as is the case when macroscopic amounts of matter produce strong magnetism. Since strong magnetism presents no problems for the Darwin Lagrangian one concludes that the problem arises in the approximation to the Legendre transform. The present author has investigated this problem in detail and at least managed to improve the situation \cite{essen96,essen97,essen99,essen&nordmark,essen06,essen08}.

It turns out that one can find very good Hamiltonians for two different limit situations. In the first case one finds the Hamiltonian as an expansion in the dimensionless parameter $N r_{\rm e}/R$, where $N$ is the number of particles with correlated velocities in a region of size $R$ and $r_{\rm e}$ is the classical electron radius. For small values of this parameter, \ie\ small density of charged particles the first couple of terms of the expansion should thus be excellent. The other limit that gives a definite result is the continuum limit of a constant, not necessarily small, density of charged particles. Below  we will consider the interaction of two systems of charged particles, one that produces a strong magnetic field, and one that responds to the energy of this field. Between them we assume that there is vacuum so the first version of the Hamiltonian, expanded to second order, should describe the interaction accurately. Note that this second order Hamiltonian, containing terms quadratic in the vector potential, is the one normally used for charged particles in an external field. Systems of many classical charged particles (plasmas) are still not well understood and it is not unlikely that the mathematical difficulties in obtaining their Hamiltonian somehow reflects this.

\section{Effective one-particle Hamiltonian}
One of the most relevant results of these investigations is a qualitatively meaningful Darwin Hamiltonian which includes quadratic terms in the Darwin vector potential (note that this is a unique object which is not subject to gauge freedom). Since quantum investigation of diamagnetism requires the square of the vector potential it seems natural that the traditional ("first order") Darwin Hamiltonian fails to describe it. Using this improved Darwin Hamiltonian \cite{essen96} one can determine the effective one particle Hamiltonian, namely all the terms that refer to, say, particle 1, in the the form \cite{essen97},
\begin{equation}\label{eq.eff.one.part.ham}
{\cal H}_1(\vecr,\vecp) = \frac{1}{2m}\left[\vecp - \frac{e}{c} \vecA(\vecr) \right]^2 +e\phi(\vecr)+V_D(\vecr)
\end{equation}
where $\phi$ is the electrostatic potential energy and,
\begin{equation}\label{eq.eff.VD.pot}
V_D(\vecr) = \frac{e^2}{mc^2}\left( \vecA\cdot\vecA_A + \frac{1}{2} \vecA_A^2 \right).
\end{equation}
Here, for our purposes,
\begin{equation}\label{eq.Adef}
\vecA(\vecr) = \frac{1}{c}\int \frac{\vecj(\vecr')}{|\vecr-\vecr'|}\dfd V' ,
\end{equation}
where the current density $\vecj(\vecr)=\varrho(\vecr)\vecv(\vecr)$ of all the other particles is the product of the charge density $\varrho$ with its velocity $\vecv$. From now on we assume that we are dealing with electrons since all effects we are discussing are much smaller for nuclei and positive ions which consequently are assumed to simply
provide a neutralizing positive background that make electrostatic effects negligible. The vector field $\vecA_A$ of Eq.\ (\ref{eq.eff.VD.pot}) is then defined by,
\begin{equation}\label{eq.A_A.def}
\vecA_A(\vecr) = \frac{e}{mc^2}\int \frac{\varrho(\vecr')\vecA(\vecr')}{|\vecr-\vecr'|}\dfd V'.
\end{equation}
In the simple symmetric cases that we will deal with here one finds that,
\begin{equation}\label{eq.A.parall.to.j}
\frac{e}{mc}\vecA(\vecr) = \nu_s \vecv(\vecr),
\end{equation}
for points where $\varrho(\vecr)\neq 0$. Here $\nu_s$ is a dimensionless constant. This gives the simple result,
\begin{equation}\label{eq.A_A.for.v.prop.A}
\vecA_A(\vecr) = \nu_s \vecA(\vecr),
\end{equation}
connecting the two vector fields. One notes that attempts to estimate $\nu_s$ in a concrete situation requires that the current density can really be seen as a product of a charge density and a velocity.

\section{The repulsive many-body force}
To find a value for $\nu_s$ we must consider a specific situation. Assume that we have a rotating superconducting sphere. On such a sphere surface current is induced that produces a magnetic field: the London moment. It is constant inside the sphere and a dipole outside so that the vector potential is given by (\ref{eq.ext.vec.pot}). Assuming that the velocity of the charge carriers is due to the rotation one can estimate the relevant surface charge density \cite{essen05} and it turns out that $\nu_s$ is order of magnitude $1$. For simplicity we use this value below, though strictly speaking it should be a bit smaller for convergence of the Hamiltonian {\em inside} the source. In most situations it is unfortunately not obvious how to find the amount of charge and corresponding velocity that produces a given $\vecj$, or $\vecB$. Nor is it clear how this is done when the magnetic field is due to ordered spins. In any case there is some reason to believe that $\nu_s$ is of order of magnitude unity in macroscopic systems. Qualitatively a small charge density and large velocity makes $\nu_s$ smaller for a given $\vecj$.

If we use cylindrical coordinates $\rho, \varphi,z$, with $r^2 = \rho^2 + z^2$, and
assume $\nu_s = 1$ the diamagnetic potential energy $V_D$ of (\ref{eq.eff.VD.pot}) is, according to (\ref{eq.ext.vec.pot}),
\begin{equation}\label{eq.VD.rot.sphere}
V_D(\vecr) = \frac{e^2}{mc^2}  \frac{3}{8} B^2  \rho^2\ \times \left\{ \begin{array}{ll} 1 & \mbox{for $0\le r \le R$} \\
R^6/r^6 & \mbox{for $R < r < \infty$} \end{array} \right. ,
\end{equation}
\ie\ harmonic attraction towards the $z$-axis for $r<R$ and a repulsion outside the sphere, except on the $z$-axis where there is no force.

Here we must carefully point out that the vector potential of the Darwin formalism has {\em no gauge freedom.} It must be divergence free, respect the symmetries of the system, have a singularity at a point charge, and go to zero far from its sources. We note that the extra (diamagnetic) term in the Hamiltonian (\ref{eq.eff.one.part.ham}) is proportional to the square of the classical electron radius $r_{\rm e}=e^2/mc^2$ since it appears also in $\vecA_A$. This type of force cannot be found simply from Maxwell's equations and the Lorentz force law, but would normally require considerations of radiation damping and the Thomson cross section. Here it arises from many body effects, in spite of neglect of radiation.

An even more interesting specific case is the field from an ideal infinite solenoid. Assuming that the radius of the cylindrical solenoid is $\rho=R$ we have that,
\begin{equation}\label{eq.A.of.solenoid}
\vecA(\vecr) = \frac{B R^2}{2\rho}\vece_{\varphi} \;\quad \mbox{for $R < \rho < \infty$},
\end{equation}
where $B$ is the constant magnetic field inside the solenoid. As is well known the field outside is zero. With our Hamiltonian (\ref{eq.eff.one.part.ham}) we thus find a repulsive force $-\partial V_D/\partial\rho \sim 1/\rho^3$, where $V_D$ is given by,
\begin{equation}\label{eq.VD.ideal.solenoid}
V_D(\vecr) = \frac{3}{2} \frac{e^2}{mc^2} \left( \frac{B  R^2}{ 2 \rho } \right)^2 ,
\end{equation}
outside the solenoid ($R<\rho$). This classical diamagnetic effect $\sim \vecA^2$ in a field free region is different from the quantum mechanical Aharonov-Bohm effect, which predicts a phase shift linear in $\vecA$. Boyer \cite{boyer73} previously investigated the classical problem but did not find the force predicted here. It is quite difficult to correctly include all contributions to the energy unless one starts from the fundamental Lagrangian and take care not to lose them. Nevertheless, an experimental test seems much needed.

Let us investigate the possibility to find the force $F_{\rho}=-\partial V_D/\partial\rho$, with $V_D$ given by (\ref{eq.VD.ideal.solenoid}) experimentally. We find that,
\begin{equation}\label{eq.Frho.ideal.solenoid}
F_{\rho}(\rho) = 6\pi r_{\rm e} R \left( \frac{B^2}{ 8\pi } \right) \left( \frac{R}{ \rho } \right)^3,
\end{equation}
where $r_{\rm e} = 2.82 \cdot 10^{-15}\,$m, and the term $B^2/8\pi$ in the first parenthesis is the magnetic energy density. In terms of SI-units this energy density is written,
\begin{equation}\label{eq.energy.dens.mag.SI}
\frac{B^2}{ 8\pi } = \frac{B^2}{ 2\mu_0 } = \half \mu_0 \left( \frac{N_t}{L} \right)^2 I^2 ,
\end{equation}
where $N_t$ is the number of turns of current $I$ in the solenoid of length $L$. The SI-unit of this is J/m$^3$. Since the classical electron radius and the radius $R$ of the solenoid both are lengths with SI-unit meter (m) we see that the unit of the force is indeed J/m = N. With $R =0.1\,$m, $B=15\,$T one finds the force $F_{\rho}(0.1\, {\rm m}) =4.76 \cdot 10^{-7}\,$N on one electron. For just $10^6$ classical electrons \footnote{A classical electron is one that has access to sufficiently many quantum states to localize in a wave packet with simultaneously defined position and velocity, see Eq. (\protect\ref{eq.mag.field.part.j}).} the force should thus be roughly half a Newton.

It is tempting to see the repulsive force discussed here as a possible explanation of the ubiquitous stellar winds emanating from magnetized plasmas. Recent work \cite{zuin&al} indicates that the parameter $N r_{\rm e}/R$, limiting the validity of our Hamiltonian, is related to the maximum electron density (Greenwald density) above which laboratory plasmas tend to disrupt. At these densities the magnetic energy per electron can become comparable to the rest energy $mc^2$ of the electrons. The fact that plasmas in general tend to be diamagnetic was noted already by Alfv{\'e}n \cite{BKalfven} and lends support to the present results.

\section{Conclusion}
The diamagnetism found here for a system in a constant magnetic field is very similar to the quantum mechanical one for a zero angular momentum atom. Since that quantum mechanical effect is closely related to the classical Larmor's theorem predicting a rotation of a system in an external field \cite{essen89} it is not really surprising that it also shows up in a classical phase space formalism that is careful to introduce canonical momenta systematically. This is after all a first step to quantum mechanics. Diamagnetism can be powerful enough to levitate macroscopic objects \cite{berry&geim} and this indicates that some form of classical understanding should be possible. Pending experimental verification of the force $-\nabla V_D$, and a full understanding of the constant $\nu_s$, our findings remain provisional and further research is desirable. But the main message here, that a Hamiltonian for a classical system of point charges that takes the Darwin velocity dependent magnetic interactions into account does predict classical diamagnetism, is hopefully now beyond dispute.\\
{\it Note added in proof:} A possible experimental verification of a classical diamagnetic current can be found in \cite{xinyong}.





\end{document}